\documentclass{jfm}

\usepackage{graphicx,
natbib,
upmath,
bm,
amsmath,
amssymb,
amsbsy,
color}
\usepackage[makeroom]{cancel}

\usepackage{algpseudocode}
\newcommand\NavSto{Navier--Stokes}

\newcommand\Rey{\mbox{\textit{Re}}}

\newcommand{\CC}{\mathrm{c.c.}}

\newcommand\ce{\mathrm{e}}
\newcommand\ci{\mathrm{i}}

\newcommand\twod{two-di\-men\-sion\-al}
\newcommand\threed{three-di\-men\-sion\-al}

\title[Estimation of unsteady aerodynamic forces] {Estimation of
  unsteady aerodynamic forces using pointwise velocity data}

\author[F. G\'omez, A. S. Sharma and H. M. Blackburn]
{F. G\'omez$^1$\thanks{Email address for
correspondence: francisco.gomez-carrasco@monash.edu},
A. S. Sharma$^2$ and  H. M. Blackburn$^{1}$}

\affiliation{$^1$Dept. of Mechanical and Aerospace Engineering, Monash
  University, VIC 3800, Australia \\[\affilskip] $^2$Aerodynamics and Flight Mechanics, \\ University of Southampton,
  Southampton SO17 1BJ, UK }

\date{?; revised ?; accepted ?. - To be entered by editorial office}

\begin{document}

\maketitle

\begin{abstract}
A novel method to estimate unsteady aerodynamic force coefficients
from pointwise velocity measurements is presented. The methodology is
based on a resolvent-based reduced-order model which requires the mean
flow to obtain physical flow structures and pointwise measurement to
calibrate their amplitudes. A computationally-affordable
time-stepping methodology to obtain resolvent modes in non-trivial flow domains
is introduced and compared to previous existing
matrix-free and matrix-forming strategies. The technique is applied to the unsteady
flow around an inclined square cylinder at low Reynolds number. The
potential of the methodology is demonstrated through good agreement
between the fluctuating pressure distribution on the cylinder and the
temporal evolution of the unsteady lift and drag coefficients
predicted by the model and those computed by direct numerical
simulation. \\[6pt] \textbf{Key words:} resolvent analysis, time-mean
flows, unsteady aerodynamic coefficients
\end{abstract}

\section{Introduction}

Unsteady motions in fluid mechanics, owing to unsteady separations and
vortex shedding, lead to unsteady aerodynamic loads of concern in
multiple engineering applications, such as flight mechanics, wind
engineering, acoustics and dynamic aeroelasticity. The identification
of unsteady aerodynamic coefficients is especially critical if new air
vehicle configurations are tested or if the flight envelope is
extended beyond traditional manoeuvres \citep*{brunton2013reduced}.
Unsteady aerodynamic models are derived either from wind tunnel
testing or directly from flight test data because unsteady simulations
of realistic configurations are likely to remain unaffordable
\citep{spalart1997comments}.  Besides classical force balance
instrumentation, non-intrusive strategies to estimate unsteady
aerodynamic forces from particle image velocimetry (PIV) are also
well-known \citep*{kurtulus2007unsteady,0957-0233-24-3-032001}. These
methods are based on combining experimental data with the governing
equations in such a way that, provided with time-resolved velocity
fields, a surface or volume integration of the \NavSto\ equations can
yield the pressure field, and hence the unsteady pressure forces. A
limitation of the methodology is that \threed\ time-resolved PIV is
required to obtain \threed\ velocity fields, and hence recover
corresponding pressure fields.

In the present work we employ a methodology to estimate unsteady
aerodynamic forces that is able to overcome the need for time-resolved
\threed\ velocity fields. Similarly to PIV-based approaches, the
present methodology is also based on the combination of measurements
with the \NavSto\ equations. However, instead of employing
\threed\ time-resolved snapshots of the velocity, the inputs of the
methodology are the time mean flow and point measurements of the
velocity. (We note that the mean flow field can notionally also be
obtained from point measurements.)
The use of the mean flow is motivated by the resolvent decomposition
of \cite{McKeonSharma2010}. A Reynolds decomposition applied to the
\NavSto\ equations reveals that the unsteady motions are dominated by
the properties of a resolvent operator depending on the mean flow and
spatial derivatives. This resolvent operator acts a
forcing-to-response transfer function at each temporal frequency,
hence the mean flow restricts the possible unsteady motions that may
exist in the flow. A singular value decomposition (SVD) of the
resolvent operator typically reveals that, at each particular
frequency, there is a dominant unsteady flow structure with
amplification ratio greater than other possible motions.  

The feasibility of employing these dominant motions as a basis for the
creation of reduced-order models of the fluctuating velocity field was
recently demonstrated by \citet{gomez_jfm1_2016} for flow in a
rectangular lid-driven cavity.
The present work expands on that theme to include the fluctuating
pressure field, hence allowing estimation of fluctuating forces on an
immersed body, and in addition employs a novel matrix-free
time-stepping algorithm to estimate resolvent SVD modes, allowing them
to be readily calculated in non-trivial flow domains.
As for the method of \citeauthor{gomez_jfm1_2016}, amplitudes of
resolvent modes used in the reduced-order model are calibrated using
pointwise measurements of the velocity.
The new methodology is applied to the estimation of fluctuating forces
imposed by the flow around an inclined square cylinder.

\section{Description of the methodology}
\label{sec.model}

\begin{figure}
\begin{center} 
\includegraphics[width=0.9\linewidth]{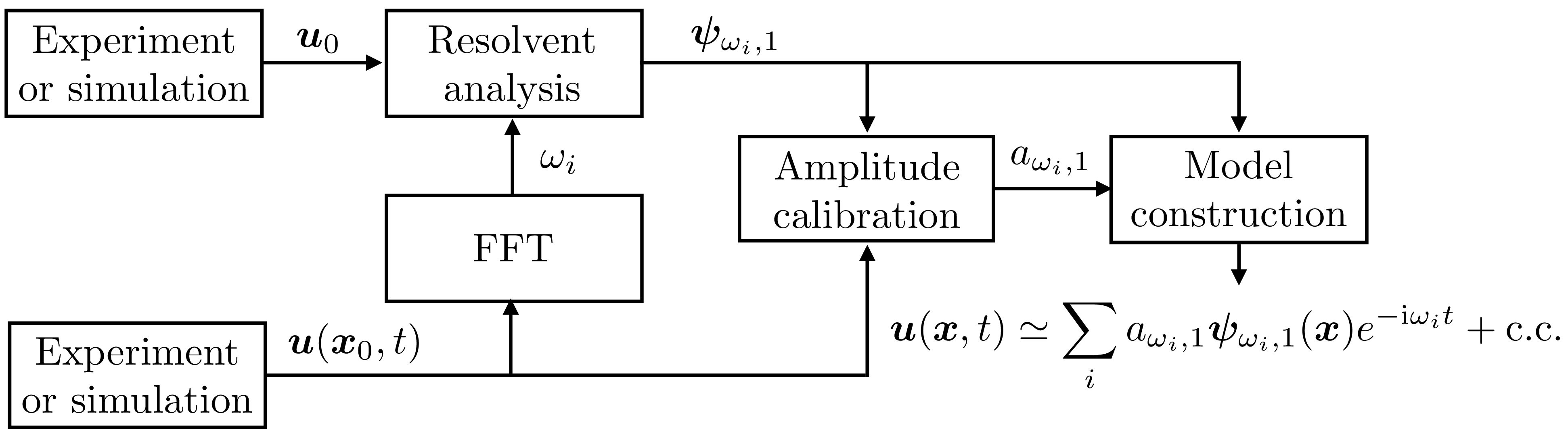} 
\end{center}
\caption{Diagram of the construction of the model. The mean flow and
  pointwise measurement inputs are on the leftmost blocks. A fast
  Fourier transform (FFT) of the probe signal provides the active
  frequencies $\omega_i$ to be explored in the resolvent analysis of
  the mean flow. The dominant resolvent modes $\psi_{\omega_i,1}$
  corresponding to the active frequencies are calibrated with the
  probe signal to obtain the amplitudes coefficients
  $a_{\omega_i,1}$. A linear combination of the weighted resolvent
  modes provides an approximation of the fluctuating velocity and
  pressure. The fluctuating pressure is included in the fluctuating
  velocity vector $\bm{u}=(u,v,w,p)^\mathrm{T}$ for convenience. }
\label{fig:diagram}
\end{figure}

Figure~\ref{fig:diagram} illustrates schematically the construction of
the model employed to estimate the unsteady forces. The time mean flow
$\bm{u}_0(\bm{x})$ and a pointwise measurement $\bm{u}(\bm{x}_0,t)$ of
the velocity history are the inputs, corresponding to the leftmost
blocks. In principle, mean flow and probe information could be
obtained independently either from experiments or simulations. A
spectral analysis of the probe signal $\bm{u}(\bm{x}_0,t)$ identifies
the active frequencies $\omega_i$ to be explored in the resolvent
analysis of the mean flow. The dominant resolvent modes
$\psi_{\omega_i,1}$ arising from the resolvent analysis corresponding
to the active frequencies are calibrated with the probe signal to
obtain the amplitudes coefficients $a_{\omega_i,1}$. A linear
combination of the weighted resolvent modes then provides an
approximation of the fluctuating velocity and associated pressure
fields.

\subsection{Resolvent decomposition}
\label{sec.resol} 
We follow a similar derivation of the resolvent decomposition as that
proposed by \cite{luhar2014opposition}. This derivation differs from
that by \cite{gomez_jfm1_2016} in that the pressure is explicitly
taken into account, instead of projecting the velocity onto a
divergence-free basis. A Reynolds decomposition is applied to the
total velocity $\bm{\hat{u}}({\bm x},t)=\bm{u}_{0}({\bm x}) +
\bm{u}({\bm x},t)$, with $\bm{u}({\bm x},t)$ being the fluctuating
velocity which may be decomposed as a sum of temporal Fourier modes
\begin{equation}
\bm{u}({\bm x},t)= \sum_{i} \bm{u}_{\omega_i}({\bm x})
\ce^{-\ci\omega_i t} + \CC 
\end{equation}
The flow is assumed to be statistically steady thus the frequencies
$\omega_i$ are real. A similar decomposition may be applied to the
nonlinear terms, leading to $\bm{f}_{\omega_i}=-(\bm{u\cdot\nabla
  u})_{\omega_i}$. These decompositions lead to a formulation of the
\NavSto\ equations as
\begin{eqnarray}
\label{eq:NSE}
 \bm{u}_{0}\bm{\cdot\nabla u}_{0}  &=& -\bm{\nabla} p +
\Rey^{-1}\nabla^2\bm{u}_{0} + \bm{f}_{0}  \\
\label{LNSE+f}
\bm{u}_{\omega_i} &=& \mathcal{H}_{\omega_i}\bm{f}_{\omega_i} \, ,
\end{eqnarray}
with $\mathcal{H}_{\omega_i}$ being the resolvent operator of the
\NavSto\ for each frequency $\omega_i$. The mean flow equation
(\ref{eq:NSE}) corresponds to $\omega=0$ and Reynolds stress
$\bm{f}_{0}$ denotes the interaction of the fluctuating velocity with
the mean. The fluctuating pressure augments the fluctuating velocity
vector as $\bm{u}=(u,v,w,p)^\mathrm{T}$ so the resolvent operator
imposes the continuity equation
\begin{equation}
\mathcal{H}_{\omega} = \left(    -\ci \omega \left[ \begin{array}{ccc}
\mathcal{I} & 0 \\
0 & 0 \\
\end{array}   \right]    -    \left[ \begin{array}{ccc}
\mathcal{L} & -\bm{\nabla} \\
\bm{\nabla}^\mathrm{T} & 0 \\
\end{array}   \right]  \right)^{-1} \left[ \begin{array}{ccc}
\mathcal{I} & 0 \\
0 & 0 \\
\end{array}   \right]\, ,
\end{equation}
with $\mathcal{L}$ being the Jacobian of the \NavSto\ equations and
$\mathcal{I}$ an identity matrix. This operator represents how the
fluctuating velocity $\bm{u}_{\omega}$ is driven by nonlinearity
$\bm{f}_{\omega}$ in Fourier space, hence it is useful to inspect its
amplification properties via a singular value decomposition (SVD)
\begin{equation}
\mathcal{H}_{\omega}=\sum_m {\bm{\psi}}_{\omega,m}
\sigma_{\omega,m} {\bm{\phi}}^*_{\omega,m} \, ,
\label{eq:SVD}
\end{equation}
where $\bm{\psi}_{\omega,m}$ and $\bm{\phi}_{\omega,m}$ are two
orthonormal basis termed response and forcing modes respectively. The
superscript $*$ indicates conjugate transpose and the subscript $m$
indicates the ordering of the modes, ranked by the amplification rate
given by the corresponding singular value $\sigma_{\omega,m}$ under
the $L_2$ energy norm. The key of the resolvent decomposition is the
projection on the nonlinearity onto the forcing modes
\citep{McKeonSharma2010}, hence the fluctuating velocity can be
written as a linear combination of response modes

\begin{equation}
\bm{u}_{\omega} = \sum_m {\bm{\psi}}_{\omega,m}
\sigma_{\omega,m}\chi_{\omega,m}\, ,
\label{eq:reseq2}
\end{equation}
where the unknown scalar coefficients $\chi_{\omega,m}$ are the
projection of nonlinearity onto forcing modes and represent the
forcing driving the velocity fluctuations \citep{gomez_jfm1_2016}.

Equation (\ref{eq:reseq2}) is an exact representation of the
\NavSto\ equation because no assumption other than a statistically
steady flow has been used. On the other hand, it is useful to exploit
the values taken by the amplification $\sigma_{\omega,m}$ in order to
construct a reduced-order model of the fluctuating velocity. In
presence of a single dominant flow feature such as a centrifugal
instability \citep{gomez_jfm1_2016} or a critical layer response
\citep{McKeonSharma2010}, the first singular value $\sigma_{\omega,1}$
is usually much larger than the second one $\sigma_{\omega,2}$, hence,
at a particular frequency $\omega_i$, irrespective of the values taken
by $\chi_{\omega,m}$, it can be assumed that the projection of
nonlinearity onto the first response ${\bm{\psi}}_{\omega,1}$ is much
larger than onto the rest. As such, the low-rank properties of the
resolvent operator can be employed to yield a rank-1 model
\begin{equation}
\bm{u}_{\omega} \simeq {\bm{\psi}}_{\omega,1} a_{\omega,1}  \, ,
\label{eq:rank1}
\end{equation}
where the amplitude coefficients
$a_{\omega,1}=\sigma_{\omega,1}\chi_{\omega,1}$ represent the amount
of nonlinearity being amplified. Under this rank-1 assumption, the
fluctuating velocity (and pressure) can be expressed as
\begin{equation}
\bm{u}({\bm x},t) \simeq \sum_{\omega} a_{\omega,1} {\bm
  \psi}_{\omega,1}({\bm x})  \ce^{-\ci\omega t} + \CC \,
\label{eq:rom} 
\end{equation}
hence this assumption provides a convenient model in which the
velocity fluctuations at each frequency are parallel to the first
singular response mode corresponding to that frequency. This rank-1
assumption has proven to be adequate in previous investigations of
pipe, channel and cavity flows \citep{McKeonSharma2010,
  Moarref2013,gomez_jfm1_2016}.

\subsection{Amplitude calibration}
\label{sec.cali}

Obtaining resolvent modes ${\bm{\psi}}_{\omega,1}$ can be
computationally challenging in complex \threed\ geometries even using
time-stepping methods. However, only a small number of modes
corresponding to the relevant or active frequencies in the flow are
computed in practice. In the absence of further information, the
active frequencies of the flow are identified via a Fourier analysis
of a pointwise measurement of the flow. As highlighted in figure
\ref{fig:diagram}, the probe information can be obtained independently
of the mean flow. Here we provide an extension of the calibration
method developed by \cite{gomez_jfm1_2016} to obtain the unknown
amplitude coefficients that close the model (\ref{eq:rom}) by using
directly the same pointwise measurements of the velocity that have
been previously employed for the identification of the active
frequencies.
At a particular spatial location $\bm{x}_0$, the reduced order model
of the fluctuating velocity satisfies
\begin{equation}
 \bm{u}({\bf x}_0, t) \simeq \sum_{i=1}^{N_{\omega}}
 {\bm{\psi}}_{\omega_i,1}(\bm{x}_0) a_{\omega_i,1} \ce^{-\ci\omega_i t}
 + \CC \, .
\label{eq:ls2}
\end{equation}
with $N_{\omega}$ representing the number of active frequencies (or
discretized frequency bins) of the flow. Although each scalar
component of (\ref{eq:ls2}) contains $N_{\omega}$ unknowns, it can be
evaluated at a number of different time instants $N_t > N_{\omega}$,
such that the solution is amenable to a least-squares
approximation. We note that the spatial structure of the fluctuating
velocity is restricted by the response modes, hence the pointwise
calibration of the amplitude coefficients serves to capture the
temporal behaviour of the fluctuating velocity. The solution of
(\ref{eq:ls2}) in a least-squares sense is given by
\begin{equation}
\mathcal{A}   = {\bm \Psi}^+{\mathcal{U}}(\bm{x_0}, t) \, 
\label{eq:ls4}
\end{equation}
with the $3 N_t\times N_\omega$ matrix ${\bm \Psi}$ containing the
values of the three velocity components of the resolvent modes and
their complex conjugates at the spatial location ${\bf x}_0$ at $N_t$
different times, the $N_\omega \times 1$ vector $\mathcal{A} $
representing the unknown amplitude coefficients, and the $3N_t \times
1$ vector ${\mathcal{U}}$ contains the values of the velocity at the
spatial location $\bm{x}_0$ at different times. The superscript $+$
denotes pseudo-inverse. The dimensions of the least-squares problem
(\ref{eq:ls4}) are much smaller than that of the SVD computations and
its solution is straightforward.

\section{Time-stepping strategies for resolvent analysis}
\label{sec:ts}

Although the simplest way to obtain numerically the resolvent modes is
to assemble the resolvent operator and perform a SVD, this is
difficult in practice owing to the massive computational requirements
resulting from the large dimensionality of the operator associated to
flows with two or three non-homogeneous spatial directions. Thus,
iterative methods are preferred.

The main idea of iterative methods is that the singular values of the
resolvent operator are the eigenvalues of
$\mathcal{H}_\omega\mathcal{H^*}_\omega$, and the response and forcing
singular vectors correspond to the eigenvectors of
$\mathcal{H}_\omega\mathcal{H^*}_\omega$ and
$\mathcal{H^*}_\omega\mathcal{H}_\omega$ respectively. As such, the
action of the resolvent operator and its conjugate transpose on a
forcing vector could enable a matrix-free iterative power method to
obtain the SVD of $\mathcal{H}_\omega$.

Following the rank-1 hypothesis posed in \S\ref{sec.resol}, only the
first singular vectors at each active frequency are required for the
construction of the reduced-order model in (\ref{eq:rom}). In this
context, \citet{monokrousos2010global} and \citet{lu2014iterative}
showed that obtaining the dominant singular vectors of the resolvent
is equivalent to finding the optimal harmonic forcing of the forced
linearized \NavSto\ equations
\begin{equation}
\label{eq:flns}
\partial_t{\bm u}({\bm x},t) = \mathcal{L} {\bm u}({\bm x},t)+ {\bm
  f}_{\omega}({\bm x}) \ce^{-\ci\omega t } \, .
\end{equation}
The long time integration of (\ref{eq:flns}) lead to the harmonic
relation (\ref{LNSE+f}), if all transient effects
vanish. Consequently, the optimal forcing and response vectors of
$\mathcal{H}_\omega$ can be obtained from an optimization problem in a
time-stepping context. A Lagrange function for the optimal forcing in
(\ref{eq:flns}) can be constructed as
\begin{multline}
{\bm L}( {\bm u}_{\omega},{\bm f}_{\omega}, {\bm v}_{\omega}, \sigma^2) =\\(  {\bm u}_{\omega},  {\bm u}_{\omega}) - \left(  {\bm v}_{\omega}, (-\ci\omega{\bm I} - \mathcal{L}^*)^{-1} {\bm u}_{\omega} - {\bm f}_{\omega} \right) - \sigma^2_{\omega} \left( ({\bm f}_{\omega}, {\bm f}_{\omega}) -1 \right) \, ,
\end{multline}
where the objective function is the fluctuation energy represented by
the energy norm $ ( {\bm u}_{\omega}, {\bm u}_{\omega})$. The first
Lagrange multiplier ${\bm v}_{\omega}$ enforces that the response and
forcing satisfy the forced linearized \NavSto\ equations
(\ref{eq:flns}) and the second Lagrange multiplier $\sigma^2_{\omega}$
enforces a unit energy norm to the optimal forcing vector. Variations
of ${\bm L}$ with respect to the fluctuating velocity $ {\bm
  u}_{\omega}$ and to the forcing $ {\bm f}_{\omega}$ yield
respectively
\begin{eqnarray}
\label{eq:res1}
{\bm v}_{\omega} &=& (-\ci\omega{\mathcal{I}} - \mathcal{L}^*)^{-1} {\bm u}_{\omega}\, , \\ 
\label{eq:res2}
{\bm f}_{\omega} &=& {\bm v}_{\omega} / \sigma^2_{\omega} \, .
\end{eqnarray}
The combination of the outcome of the optimization problem (\ref{eq:res1}) and (\ref{eq:res2}) with the (\ref{LNSE+f}) leads to the eigenvalue problem
\begin{equation}
\label{eq:evp2}
{\bm \psi}_{\omega,1} =\sigma^{-2}_{\omega,1} (-\ci\omega{\mathcal{I}} - \mathcal{L})^{-1}(\ci\omega{\mathcal{I}}  - \mathcal{L}^*)^{-1}{\bm \psi}_{\omega,1} \, .
\end{equation}
where the fact that the optimal forcing solution corresponds to the
first (most amplified) forcing mode ${\bm \psi}_{\omega,1} $ has been
exploited. An iterative matrix-free power method can be then applied
to the eigenvalue problem (\ref{eq:evp2}) in order to obtain the
dominant eigenvalue. However, as described by
\citet{monokrousos2010global}, time-stepping does not directly provide
a solution of (\ref{LNSE+f}) or (\ref{eq:res1}) and the forced
linearized and adjoint equations need to be integrated long enough
such that the transient dynamics vanish and the response is
harmonic. Note that the long integration of the adjoint of
(\ref{eq:flns})
\begin{equation}
\label{eq:aflns}
\partial_t{\bm v}({\bm x},t) = \mathcal{L}^* {\bm u}({\bm x},t)+  {\bm f}_{\omega}({\bm x}) \ce^{\ci\omega t } \, ,
\end{equation}
provides a harmonic solution for (\ref{eq:res1}).

Although the algorithm originally proposed by
\citet{monokrousos2010global} is successful in the sense that it is
matrix-free, a number of difficulties arise in practice. The time
integration required of (\ref{eq:flns}) until a harmonic response is
obtained can be very long, hence the method may not be computationally
affordable.  The method is particularly slow at low frequencies. Also,
the complex velocity response ${\bm u}_{\omega} $ is obtained via a
Fourier transform of ${\bm u}(t)$ during one period, hence it could
have a different phase at each iteration step. Finally, the response
of (\ref{eq:flns}) can be susceptible to large transient growths
depending on the initial condition. Even if ${\bm u}_{\omega} $ and
${\bm f}_{\omega} $ are close to convergence, the transient can be
significant if the two vectors do not have their relative phase fixed
by the resolvent.

A modification of the time-stepping approach of
\citet{monokrousos2010global} which is able to cope with the
above-mentioned limitations is proposed by allowing the velocity field
to take on complex values into the forced direct and adjoint
linearized \NavSto\ equations. The use of a complex variable has the
advantage that is it is not necessary to perform a Fourier transform
of the response during one period to obtain ${\bm u}_{\omega}$. Once
the response of (\ref{eq:flns}) is harmonic, any snapshot of the
complex vector ${\bm u}^\prime({\bm x},t)$ represents its Fourier
transform ${\bm u}_{\omega} $ at some particular phase. Furthermore,
if the time integration is taken as an integer number $n$ of periods
$T=2\pi/\omega$ the response and force vectors obtained remain with
the same relative phase imposed by the resolvent during a complete
iteration. This feature avoids possible sources of transient effects
during iterations.

In addition, the integration time can be limited to a small number of
periods $n$ such that the strongest transients are vanished but the
flow is not yet exactly periodic. This permits obtaining an estimation
of the amplification value and of the response and forcing vector in
relatively short integrals. The estimated $\sigma$ is then employed
between each direct and adjoint iteration to generate a initial
condition for the forced equations~(\ref{eq:flns}). This initial
condition can be understood as a preconditioner, and it approaches the
correct response as the estimated value $\sigma$ gets closer to the
exact value. As a result, the transient effects are damped and the
harmonic response is achieved after a few iterations.

All these features permit obtaining the most amplified modes using a
smaller computational effort than with the original algorithm of
\citet{monokrousos2010global}. The modified algorithm is listed in
table~\ref{tab:algorithm}. In practice, the harmonic response is also
assessed by monitoring the fluctuation energy. If the difference
between the maximum and minimum fluctuation energy is less than a
given tolerance within one period, the response is considered harmonic
and the integration stops. Thus each iteration may require a number of
integration periods less than $n$.

\begin{table}
    \begin{algorithmic}[1]          \State Set a random initial unit norm forcing ${\bm f}^0_{\omega}$ and amplification $\sigma^0_{\omega}$
         \While{$ | {\bm f}^{i+1}_{\omega} - {\bm f}^{i}_{\omega} | $ is larger than a given tolerance}
            \State Set initial condition ${\bm u}^i_{\omega}(0)=\sigma^i_{\omega}{\bm f}^i_{\omega}$ and integrate  (\ref{eq:flns}) over $nT$ to obtain ${\bm u}^{i+1/2}_{\omega}={\bm u}^i_{\omega}(nT)$
                \State Update amplification $\sigma^{i+1/2}_{\omega} =  \| {\bm u}^{i+1/2}_{\omega} \| $ and normalize  ${\bm u}^{i+1}_{\omega} = {\bm u}^{i+1/2}_{\omega} / \sigma^{i+1/2}_{\omega} $
                \State  Set initial condition ${\bm v}^{i}_{\omega}(0)=\sigma^{i+1/2}_{\omega}{\bm u}^{i+1}_{\omega}$ and integrate  (\ref{eq:flns}) over $-nT$ to obtain ${\bm v}^{i+1/2}_{\omega}={\bm v}^i_{\omega}(-nT)$. 
                 \State Update amplification $\sigma^{i+1}_{\omega} =  \| {\bm v}^{i+1/2}_{\omega} \| $ and set the new {forcing}  ${\bm f}^{i+1}_{\omega} = {\bm v}^{i+1/2}_{\omega} / \sigma^{i+1}_{\omega} $.
            \EndWhile
 \caption{\label{tab:algorithm} Modification of the algorithm proposed by \citet{monokrousos2010global}.  }
    \end{algorithmic}
\end{table}

\begin{figure}
\begin{center} 
\includegraphics[width=1\linewidth]{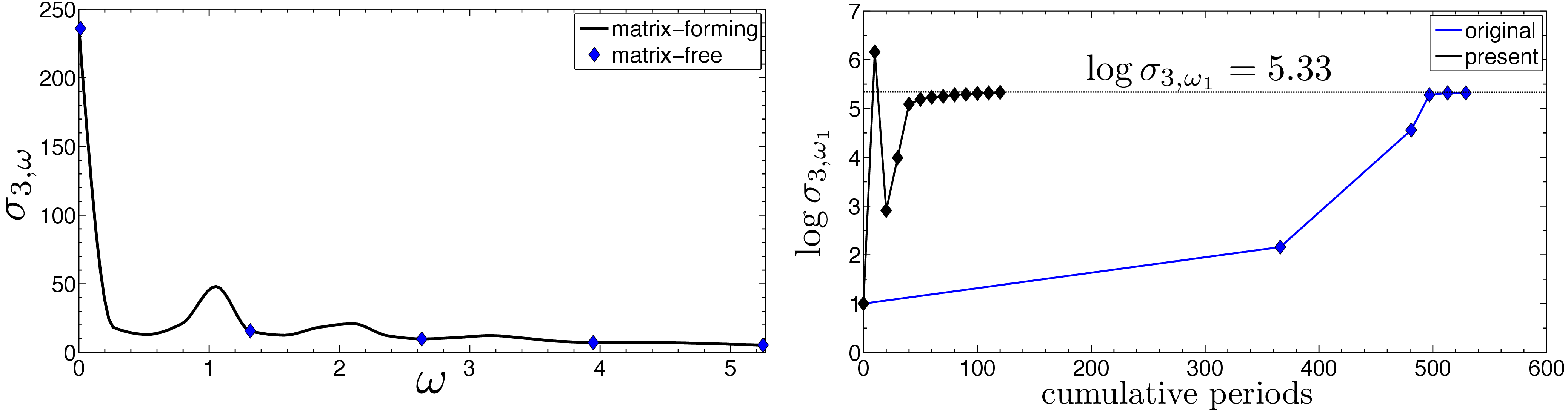} 
\end{center}
\caption{ (left) Validation of the method: amplitude of the first
  resolvent modes in frequency at $\beta=3$ obtained with the
  matrix-forming and time-stepping matrix-free methods. (right)
  Comparison of the original time-stepping methods envisaged by
  \citet{monokrousos2010global} and the present approach using $n=5$:
  cumulative integration periods until convergence of the
  amplification. Diamond symbols denote the estimated value at each
  iteration. $\beta=3$, $\omega_1=5.26$. The tolerance error is set to
  $0.005$ and the two methods start with the same initial condition.}
\label{fig:val}
\end{figure}

\subsection{Validation and comparison of the method}

A validation of the computation of the resolvent modes via
time-stepping is carried out against results obtained using an
in-house matrix-forming shift-invert method.  The chosen validation
case is the flow enclosed in a square lid-driven cavity at $Re=1200$
with a periodic span $\Lambda=0.945$ \citep{gomez_jfm1_2016}. A good
agreement between the two methods is observed in
figure~\ref{fig:val}(a). However, for the problem examined, while the
matrix-forming approach requires the storage and evaluation of a
$\mathcal{O}\left(N^2\right)$ matrix, the time-stepping approach only
needs two planes with $\mathcal{O}\left(N\right)$ degrees of freedom
to deliver the same results. In the case presented, this evaluates to
$3 \times 64^2$ against $(3 \times 64^2)^2$.

The optimal value of the parameter $n$ depends on the problem and, for
the present case, we have observed that the method typically converges
in a few iterations using $n=5$. In other words, five periods are
enough to damp the most significant transient effects in the present
case. Figure~\ref{fig:val}(b) shows a comparison of the present method
with the original time-stepping methods envisaged by
\citet{monokrousos2010global}. We observe that although the proposed
methods requires a larger number of iterations, it needs significantly
less integration periods to achieve convergence. In this example, the
number of the required integration periods until convergence goes from
540 periods to 116 periods, which translates into a $\sim80\%$ saving
of CPU time.

\section{Application to the flow around an inclined square cylinder}

The flow past a \twod\ inclined square cylinder may serve as a model
for the flow around a non-trivial bluff body and it is a good
compromise between computational affordability and complex flow
features \citep{sohankar1998low,yoon2010flow}.  Beyond the critical
Reynolds number, the wake becomes unsteady presenting asymmetric
vortex shedding, thus this flow is interesting for the investigation
of unsteady lift and drag forces.

Although the present methodology seems more appealing for experimental
works, the mean flow and pointwise measure inputs to the model have
been obtained via direct numerical simulation (DNS) using a
spectral-element solver \citep{hugh2002}. A rectangular computational
domain defined in $[-16,20]\times[-14,14]$ has been discretized with
236 spectral elements. The square cylinder has a unit side length and
its centroid is located at $(x,y)=(0,0)$. Temporal and spatial
convergence has been achieved with a polynomial expansion of order 11
in each element and using a second-order temporal scheme with $\Delta
t = 8.5 \times 10^{-3}$. A constant velocity
$(u,v)=(\cos{\alpha},\sin{\alpha})$ is imposed at the inlet of the
domain, no-slip boundary condition at the cylinder wall and Neumann
boundary conditions are imposed at the outlet. The Reynolds number
based on the cylinder side length $D$ and the modulus of the inlet
velocity is fixed to $Re=100$. The angle of attack is set to
$\alpha=10^\circ$.

\begin{figure}
\begin{center} 
\includegraphics[width=0.9\linewidth]{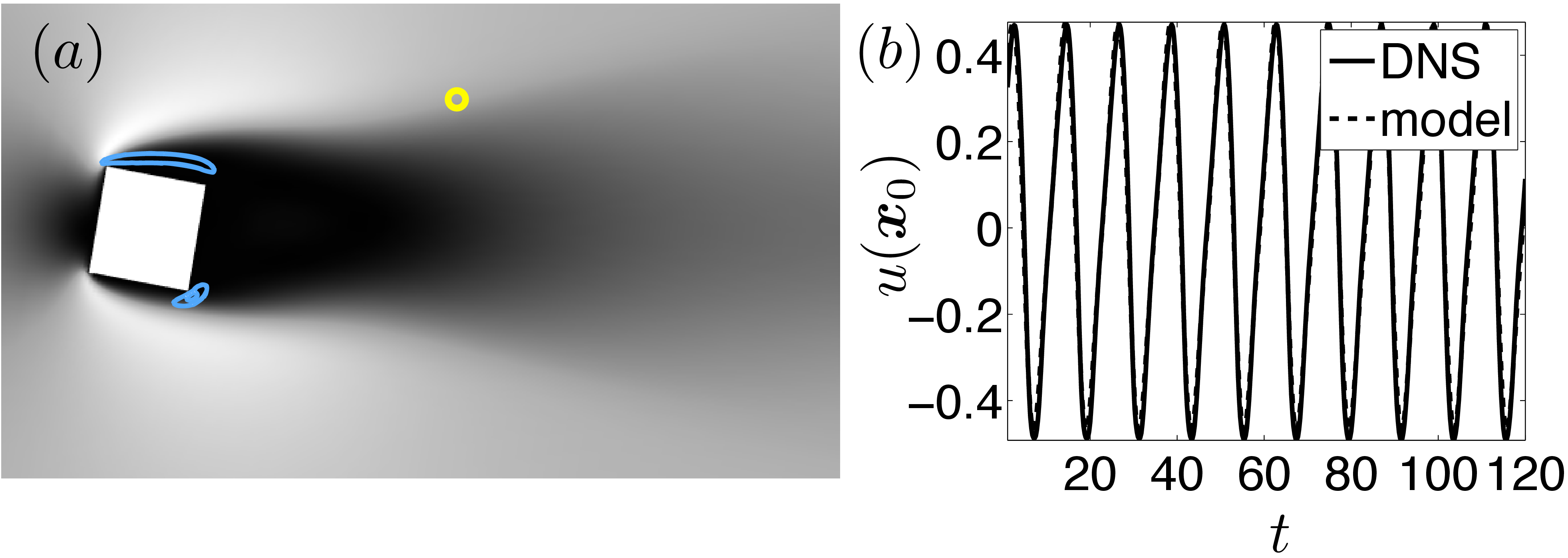} 
\end{center}
\caption{Inputs to the model corresponding to the leftmost block in
  figure~\ref{fig:diagram} (a) Streamwise velocity contours of the
  mean flow, colored from black to white. Blue contour lines indicate
  40\% and 80\% of the maximum of the kinetic energy corresponding to
  $\bm{\phi}_{\omega_1,1}$.(b) Evolution of the fluctuating streamwise
  velocity in the wake at the location $\bm{x}_0=(3,1.5)$, measured
  from the DNS and fitted to the model. The probe position is
  highlighted in (a) with a yellow circle.}
\label{fig:input}
\end{figure}

The inputs to the model corresponding to the leftmost block in
figure~\ref{fig:diagram} obtained via DNS, mean flow and a single
pointwise measurement of the velocity, are shown in
figure~\ref{fig:input}. The probe is located at a random point in the
wake of the cylinder where the shear is non-zero and it serves to
identify a single dominant frequency $\omega_1 = 0.908$, hence only
one resolvent mode corresponding to that frequency needs to be
computed. The dynamics of the self-interaction of this mode in this
kind of flows could be relevant \citep{Noack2003jfm} thus it may be
convenient to also consider the first harmonic $\omega_2 = 2
\omega_1$.

The iterative time-stepping algorithm described in \S\,\ref{sec:ts} has
been employed to obtain the two resolvent modes required to construct
the reduced-order model. The forced linearized \NavSto\ equations and
their adjoint version have been solved with the same spectral-element
solver employed for DNS. The boundary conditions for each equations
are described by \citet{barkley2008direct}, however two set of
boundary conditions at the inlet have been tested the forced adjoint
equations, (i) an extended domain $[-40,20]\times[-14,14]$ with
Dirichlet boundary conditions and (ii) the same domain employed for
the DNS with a non-physical forcing $-m(\bm{x})\bm{u}(\bm{x},t)$
applied at $-16<x<12$ to force a zero-amplitude of the forcing mode at
$x=-16$. Both boundary conditions have provided similar results, thus
the former has been adopted on account of smaller computational
requirements.

\begin{figure}
\begin{center} 
\includegraphics[width=0.45\linewidth]{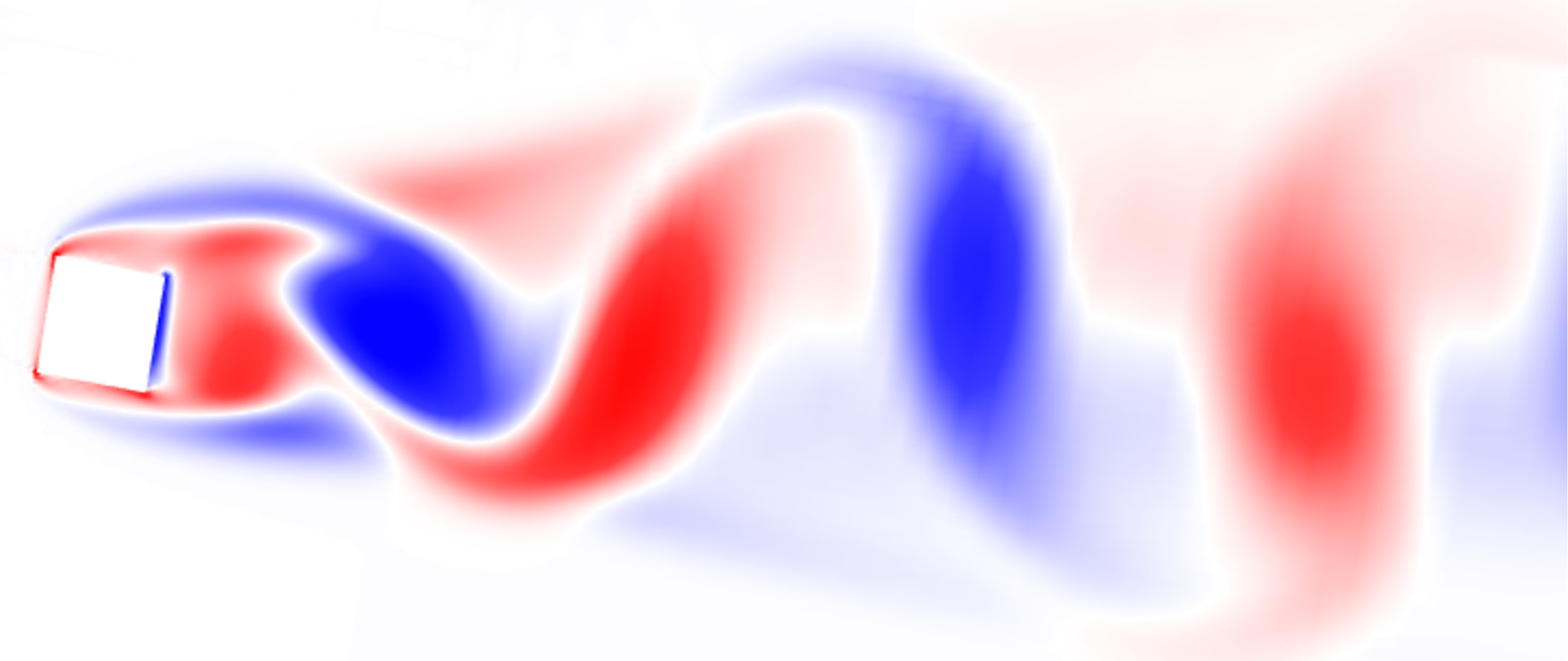} 
\includegraphics[width=0.45\linewidth]{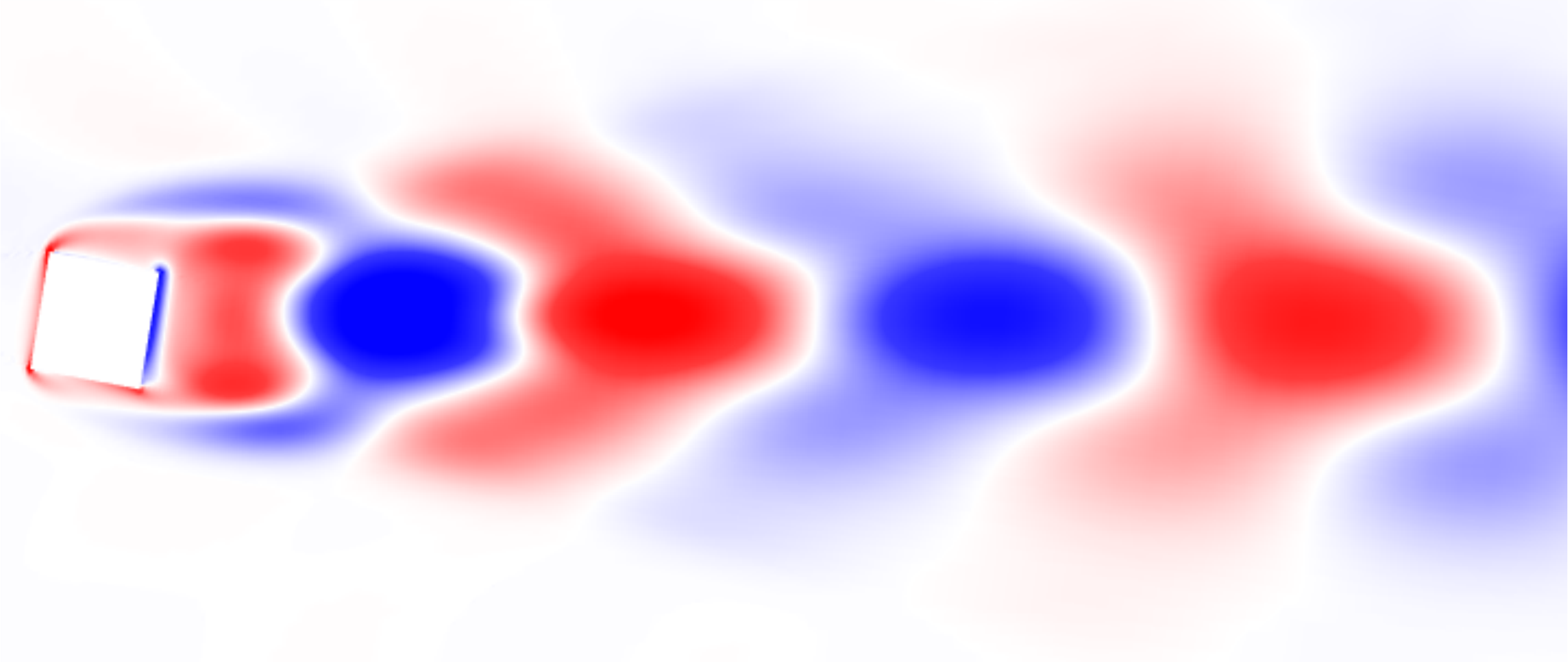} 
\end{center}
\caption{Comparison of the vorticity fields obtained from (left) DNS
  and (right) resolvent-based model via calibration of the amplitude
  against the probe data in figure~\ref{fig:input}(b). Colored
  contours represent $\pm 1/3$ of the maximum and minimum vorticity}
     \label{fig:vort}
\end{figure}

\begin{figure}
\begin{center} 
\includegraphics[width=1.0\linewidth]{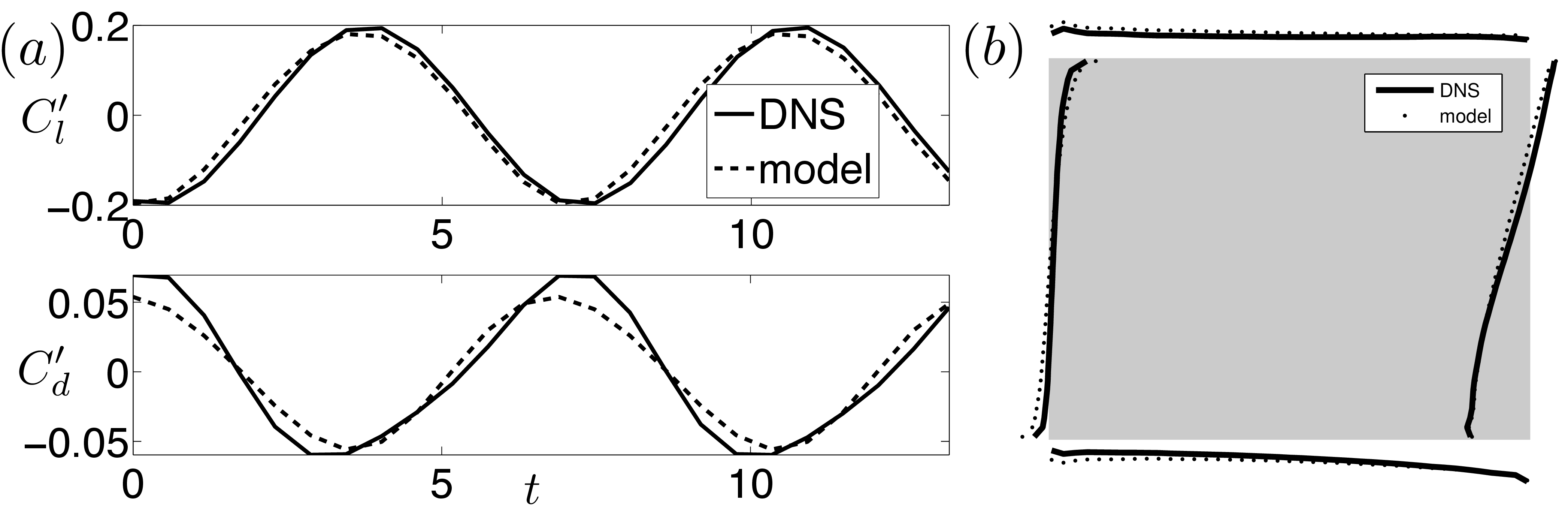} 
\end{center}
\caption{ (a) Comparison of the temporal evolution of the unsteady
  lift and drag coefficients calculated via DNS and predicted by the
  present methodology. (b) Representation of the fluctuating pressure
  distribution along the sides of the cylinder at a random instant
  calculated via DNS and predicted by the present methodology. Each
  cylinder side acts as a $x-$axis while their normal direction
  indicates the relative value of unsteady pressure at that location.}
\label{fig:forces}
\end{figure}

A comparison of the vorticity fields obtained from DNS and the
resolvent-based model via calibration of the amplitude against the
probe data in figure~\ref{fig:input}(b) is shown is
figure~\ref{fig:vort}. It is remarkable that flow in the region around
the cylinder, and hence the unsteady separation, is accurately
predicted by the model, despite the probe being located approximately
three side lengths from the cylinder. On the other hand, the structure
of the wake far from the cylinder present obvious
discrepancies. Although the general feature of vortex shedding is also
reproduced, the DNS presents additional features that the present
resolvent-based mode does not capture, like a consecutive and opposite
vertical displacement of the vortex cores. An inspection of the
spatial structure of the forcing mode that drives the resolvent modes,
shown in blue isolines in figure~\ref{fig:input}(a), reveals that
their maxima are located within the boundary layer of the cylinder,
thus the discrepancies between the model and DNS can be attributed to
this observation. As such, we presume that the additional dynamics of
the far wake are governed by additional subdominant resolvent modes
associated with the shear in the wake not considered in the present
model.

The present approach is validated by measuring the unsteady lift and
drag coefficients defined as:
\begin{equation}
C^\prime_l = \int_{\delta\Omega_c} p(\bm{x}) \bm{n} \cdot \bm{e}_y, \qquad
C^\prime_d = \int_{\delta\Omega_c} p(\bm{x}) \bm{n} \cdot \bm{e}_x \, .
\end{equation}
where ${\delta\Omega_c}$ denotes the boundary of the cylinder,
$\bm{n}$ is the normal vector around the square cylinder while
$\bm{e}_x$ and $\bm{e}_y$ are the unit vectors in the streamwise and
normal direction, respectively. At the present value of $Re$, the
contributions of the viscous forces are negligible. Nevertheless, they
could be taken into account using the present methodology. A good
agreement between the temporal evolution of the unsteady lift and drag
coefficients calculated via DNS and predicted by the present
methodology is shown in figure~\ref{fig:forces}(a). This results is
consistent with the accurate prediction by the model of the near field
around the cylinder shown in figure~\ref{fig:vort}. To provide further
insight, the fluctuating pressure distribution along the sides of the
cylinder at a random instant is represented in
figure~\ref{fig:forces}(b). The resemblance between the pressure
distributions corresponding to DNS and the resolvent-based model
supports the good agreement between the temporal evolution of the
unsteady lift and drag forces.  Finally, the same approach has been
carried out using different pointwise measurements. As remarked by
\citet{gomez_jfm1_2016}, similar results have been obtained provided
that the probe is always positioned at a location where the
fluctuating velocity is significant. However, this is not an issue
because the locations where the fluctuating velocity is significant
can be inferred from the spatial structure of the resolvent modes.

\section{Conclusions}
\label{sec:con}
A novel method to estimate unsteady aerodynamic coefficients via
pointwise measurements has been presented. The methodology requires
two inputs (i) the mean flow and (ii) temporal information from a
probe. In principle, both inputs could be obtained either
simultaneously or independently. Although we believe the present
methodology is more appealing for experimental investigations,
e.g. using planar time-resolved PIV to obtain a \threed\ mean flow and
obtain temporal information at different locations, DNS was employed
in the present work to obtain the mean flow and the pointwise data.

The most challenging step of the methodology is the computation of the
resolvent modes. For this purpose, a computationally-affordable
time-stepping methodology to obtain resolvent modes of complex flows
has been introduced, validated and compared to previous existing
matrix-free and matrix-forming strategies.

The potential of the present methodology has been demonstrated by
application to an unsteady \twod\ flow around an inclined square
cylinder at low Reynolds number. The present approach can predict the
fluctuating velocity associated with unsteady separation and the
pressure distribution near the square cylinder using just the leading
response mode. The temporal evolution of lift and drag coefficients
computed from those fields are in good agreement with those obtained
via DNS.

\bibliography{cylinder}
\bibliographystyle{jfm}

\end{document}